\documentclass[acmsmall,screen=True,natbib=True,review=False,anonymous=False]{acmart}
\usepackage[csdisplay=true,splitcomp=true,thresholdtype=words,threshold=15,autopunct=true]{csquotes}

\usepackage{longtable}

\usepackage{color, colortbl}
\definecolor{lavenderblue}{rgb}{0.9, 0.9, 0.98}
\usepackage{graphicx}
\usepackage{multirow}
\usepackage{svg}
\usepackage{afterpage}

\usepackage{etoolbox}
\makeatletter
\pretocmd{\NAT@citexnum}{\@ifnum{\NAT@ctype>\z@}{\let\NAT@hyper@\relax}{}}{}{}
\makeatother

\def\citepos#1{\citeauthor{#1}'s \citep{#1}}

\AtBeginDocument{\providecommand\BibTeX{{\normalfont B\kern-0.5em{\scshape i\kern-0.25em b}\kern-0.8em\TeX}}}

\setcopyright{acmcopyright}
\copyrightyear{2018}
\acmYear{2018}
\acmDOI{10.1145/1122445.1122456}

\acmConference[Woodstock '18]{Woodstock '18: ACM Symposium on Neural
  Gaze Detection}{June 03--05, 2018}{Woodstock, NY}
\acmBooktitle{Woodstock '18: ACM Symposium on Neural Gaze Detection,
  June 03--05, 2018, Woodstock, NY}
\acmPrice{15.00}
\acmISBN{978-1-4503-XXXX-X/18/06}

\acmSubmissionID{2272}

\usepackage{xcolor}
\usepackage{tikz}
\usepackage[htt]{hyphenat}
\def\Slash{\slash\hspace{0pt}}

\begin{document}

\title[No Community Can Do Everything]{No Community Can Do Everything: Why People Participate in Similar Online Communities}

\author{Nathan TeBlunthuis}
\affiliation{\institution{University of Washington}
  \city{Seattle}
  \country{USA}}
    \orcid{0000-0002-3333-5013}
\email{nathante@uw.edu}

\author{Charles Kiene}
\affiliation{\institution{University of Washington}
  \city{Seattle}
  \country{USA}}
  \orcid{0000-0002-7327-8391}
\email{ckiene@uw.edu}

\author{Isabella Brown}
\affiliation{\institution{University of Washington}
  \city{Seattle}
  \country{USA}}
  \orcid{0000-0002-6790-1976}
\email{isabrown@uw.edu}

\author{Laura (Alia) Levi}
\affiliation{\institution{University of Washington}
  \city{Seattle}
  \country{USA}}
  \orcid{0000-0003-0384-7488}
\email{@uw.edu}

\author{Nicole McGinnis}
\affiliation{\institution{University of Washington}
  \city{Seattle}
  \country{USA}}
    \orcid{0000-0001-7638-220X}
\email{@uw.edu}

\author{Benjamin Mako Hill}
\affiliation{\institution{University of Washington}
  \city{Seattle}
  \country{USA}}
  \orcid{0000-0001-8588-7429}
\email{makohill@uw.edu}

\renewcommand{\shortauthors}{TeBlunthuis, et al.}

\begin{abstract}

Large-scale quantitative analyses have shown that individuals frequently talk to each other about similar things in different online spaces. Why do these overlapping communities exist? We provide an answer grounded in the analysis of 20 interviews with active participants in clusters of highly related subreddits. Within a broad topical area, there are a diversity of benefits an online community can confer. These include (a) specific information and discussion, (b) socialization with similar others, and (c) attention from the largest possible audience. A single community cannot meet all three needs. 
Our findings suggest that topical areas within an online community platform tend to become populated by groups of specialized communities with diverse sizes, topical boundaries, and rules. Compared with any single community, such systems of overlapping communities are able to provide a greater range of benefits.

\end{abstract}



\maketitle

% \fontsize{12pt}{24pt}
% \selectfont

\section{Introduction}

Early work in social computing treated online communities as isolated units that could be understood without considering their members' participation in other online communities. As community hosting platforms such as Reddit and Facebook have grown in prominence, social computing scholars have sought to document and explore the connections between online communities \citep{datta_extracting_2019, hill_studying_2019, tan_all_2015, zhu_selecting_2014}.
This research has shown that online communities overlap with each other in terms of their memberships and topics in ways that have important consequences for a range of outcomes \cite{teblunthuis_identifying_2021, chandrasekharan_internets_2018, wang_impact_2012}.

User and topic overlap is widespread---both within platforms and across them. 
For example, a range of studies have highlighted the fact that members frequently participate in multiple online groups. 
This occurs both serially as users migrate between communities over time \cite{lu_investigate_2019, tan_all_2015, tan_tracing_2018} and concurrently as individuals belong to multiple groups at once \citep{wang_impact_2012, hwang_why_2021, zhu_impact_2014}. 
Many large platforms host distinct communities with similar topics and content \citep{datta_identifying_2017, zhu_selecting_2014}.
In at least one study, researchers have documented that overlaps in users and topics often coincide \cite{datta_identifying_2017}.
In other words, members of online communities often simultaneously participate in overlapping conversations with overlapping groups of people in different online spaces. 

\textit{Why are the same individuals talking to each other about similar things in different online communities?}
Although social computing offers many theories of why individuals might want to participate in a community, almost all empirical work in social computing on user and topic overlap has used computational or quantitative analysis. As a result, we know very little about what overlaps mean to users. Critically, we also have very little in the way of empirical evidence that is able to speak to why communities overlap in the first place.

Our work seeks to complement existing quantitative research with a better qualitative understanding of intercommunity overlap and contribute to several streams of social computing scholarship. 
In particular, our work complements a series of social computing studies that have taken inspiration from ecological theory and shown that online groups' growth and survival are closely tied to activity in adjacent online spaces \cite{teblunthuis_identifying_2021, wang_impact_2012, zhu_impact_2014, zhu_selecting_2014}.

We seek to answer our research question (in italics above) through an interview-based study of Reddit users with experience in overlapping communities. Using a dataset of posts and comments on Reddit, we identify clusters of communities on Reddit with highly overlapping users and topics and recruit a set of 20 participants from nine clusters.
Drawing from a grounded theory analysis of interview transcripts, we develop an explanation of why many users participate simultaneously in communities with overlapping memberships and topics. 

Our findings suggest that users seek three benefits from online groups: users want to (a) find specific types of content, discussions, and information; (b) connect with similar types of people; and (c) share content with the largest possible audience. Our work also suggests that these three benefits are frequently in conflict such that the more a community provides one of these benefits, the less able it will be to provide the other two. Because it is difficult for a single community to fully provide all three benefits, clusters of multiple overlapping communities are constructed to do so in aggregate.

\section{Related Work}

\label{sec:overlapping}

Although most research in online communities analyzes the internal factors driving online community success \cite{kraut_building_2012}, a growing literature studies communities related by overlaps in topic or membership \cite{datta_identifying_2017, tan_all_2015, teblunthuis_identifying_2021, zhu_impact_2014}. 
This work has found that concurrent engagement in multiple communities is common on large platforms that host online communities such as Reddit where individuals smoothly jump from community to community \cite{tan_all_2015}.  
With several exceptions \cite[e.g.,][]{fiesler_moving_2020, kiene_technological_2019,  zhao_social_2016, hwang_why_2021}, this work typically takes the fact that communities overlap for granted and focuses on the consequences of overlap on outcomes such as the emergence and growth of communities \cite{butler_cross-purposes_2011, zhu_impact_2014} and the diffusion of types of language such as hate speech \cite{chandrasekharan_internets_2018}. None of this work provides insight into how communities come to overlap and why these overlaps persist.

Researchers have investigated intercommunity conflict and found that conflict is initiated by a very small proportion of online communities \cite{kumar_community_2018}.  
Other work has shown that content cross-posted to different communities contributes to the ongoing renegotiation of topical boundaries \cite{butler_cross-purposes_2011}. 
\citet{zhang_understanding_2021} have shown that topical boundaries can also shift as similar communities attract users with different interests.
In a related sense, \citet{massanari_gamergate_2017} has argued that toxic communities can influence the broader culture of a platform. As a result, banning problematic communities from a platform such as Reddit can reduce toxicity in adjacent communities that are not directly affected \citep{chandrasekharan_you_2017, ribeiro_platform_2021}.

A number of studies on overlapping communities draw upon ecological theory \cite{teblunthuis_identifying_2021, wang_impact_2012,  zhu_impact_2014, zhu_selecting_2014}.
Ecological approaches in social computing theorize that overlaps between users and topics relate to competitive or mutualistic forces and drive outcomes such as growth and survival. For example, \citet{wang_impact_2012} found that membership overlap reduced the growth rate of Usenet groups. \citet{zhu_selecting_2014} found that participation rates often suffered if there was too little or too much overlap with other communities. 
\citet{zhu_impact_2014} found that communities' survival was positively associated with membership overlap, especially with overlap with older communities. Recently, \citet{teblunthuis_identifying_2021} found that mutualism is common in clusters of overlapping subreddits.

Although these studies use statistical analysis to tell us about how communities relate to each other, they do not to speak to \textit{how} participants understand the relationships between similar online communities or \textit{why} they participate in overlapping communities. The exclusively quantitative nature of these accounts means that a range of potential explanations are possible.

Although we know of no qualitative examination focused directly on understanding why overlapping communities exist, there are a series of qualitative papers that point to potential answers. \citet{fiesler_moving_2020} describe the history of online fanfiction writing communities migrating across platforms in pursuit of hospitable infrastructure. Similarly, \citet{zhao_social_2016} describe how individuals use multiple social media platforms to meet varied and nuanced communication needs.
Although their study is primarily quantitative, \citet{zhu_selecting_2014} include quotes from interviews to support the emic validity of notions of competition and mutualism between groups in an enterprise social media system.  
Finally, \citepos{hwang_why_2021} recent paper seeks to explain why individuals participate in persistently small online communities on Reddit and ends with a reflection that many small communities are sustainable only because they are ``nested'' within larger niches.
All told, these findings suggest a rich social process by which participants in online communities purposefully construct and move between overlapping spaces.

% Social computing scholarship has pointed to differences in affordances between platforms and the ways in which members migrate between communities over time.
However, the small amount of qualitative evidence from participants in overlapping communities means that we lack a strong sense of why members choose to participate in multiple communities simultaneously. 
Although ecological studies attempt to quantify competition and mutualism, we know little about how members understand the relationships between their communities or if these key ecological concepts have any emic resonance.  
Our work seeks to place ecological studies of online communities on firmer qualitative ground.

\subsection{Reasons for Joining Online Communities}
\label{sec:reasons}

Decades of social computing research has sought to understand why people belong to particular online communities \cite{kraut_building_2012}. 
It has long been recognized that different people have different motivations and that a single individual may have multiple motivations that include the social, informational, and material benefits users receive through their participation \cite{butler_membership_2001,  turner_where_2005, xigen_li_factors_2011}. In terms of uses and gratifications theory, ``users actively seek particular media with the goal of gratifying an existing need'' \cite{lampe_motivations_2010}. 
Past research has shown that people seek online communities to collaborate on projects \cite{poor_computer_2014}, to receive social support \cite{leimeister_evaluation_2005}, to cooperate with friends \cite{turner_where_2005}, and, especially, to exchange information \cite{ridings_virtual_2004, muhtaseb_arab_2008, leavitt_role_2017, liang_knowledge_2017}. 
Other research focuses on the growth and decline of membership in online communities and surfaces motivations for why people choose not to participate \citep{cunha_are_2019}. \citet{brandtzaeg_user_2008} found that a lack of trust or low quality content can lead to declines in membership.  
Online communities may decline because leaders are resistant to change and unwelcoming to newcomers \citep{shaw_laboratories_2014, halfaker_rise_2013, teblunthuis_revisiting_2018}.

Although our findings are the result of an inductive process of bottom-up grounded theory analysis, the presentation of our findings relies on three existing concepts: finding specific content, homophily, and finding the largest possible audience.

\subsubsection{Finding specific content}

One of the most important features of online communities is their ability to enable the spread of useful knowledge and information \cite{faraj_online_2016}. By connecting individuals with specific information and skills that they desire, online communities match knowledge seekers with experts and foster collaboration on information goods \cite{benkler_wealth_2006,  lakhani_how_2003, fulk_connective_1996,fiesler_growing_2017}.
Research has often focused on the ways that individuals utilize diverse types of social computing systems to meet their specific information needs through systems such as Q\&A sites \cite{adamic_knowledge_2008}, synchronous chat systems \cite{white_effects_2011}, search engines \cite{morris_comparison_2010}, social network sites \cite{starbird_crowd_2012,morris_what_2010}, fanworks \cite{fiesler_growing_2017}, and knowledge bases \cite{ackerman_answer_1990,orlikowski_learning_1992}.  

\subsubsection{Homophily}

A second need that online communities serve is to foster connections with similar others. The term \textit{homophily}, ``a tendency for friendships to form between those who are alike in a designated respect''  \cite{lazarsfeld_friendship_1954}, describes the set of benefits people can only receive from others who share their identities, beliefs, interests, or culture \cite{mcpherson_birds_2001}.  
In offline settings, homophily helps explain why tastes in cuisine, music, and other cultural preferences are often correlated \cite{dellaposta_why_2015}, why similar people tend to congregate, and what happens when they do \cite{mcpherson_birds_2001}. 
Homophily on social networks may drive the emergence of online ``echo chambers'' as individuals seek online communities whose members share their beliefs \cite{johnson_communication_2009, grevet_managing_2014, himelboim_valence-based_2016, dvir-gvirsman_media_2017}. 

Research has shown that people have greater degrees of trust in homophilous groups and are more likely to share content posted by homophilous others \citep{ma_when_2019, chang_specialization_2014}.
Homophily has been described as an important feature of online fan communities \citep{hillman_alksjdflksfd_2014,fiesler_growing_2017}.

\subsubsection{Finding the largest possible audience}

Research on online communities producing public information goods has found evidence that audience size motivates contributors \cite{zhang_group_2011}. Additionally, numerous studies have shown that users of social networking sites frequently consider the audience that their posts and messages may reach \cite{marwick_i_2011, zhang_configuring_2020}. 
As individuals on social media typically have little information about who sees their posts, they conceive of ``imagined audiences'' based on cues from visible activity \citep{bernstein_quantifying_2013} and target imagined audiences using deliberate strategies, such as using multiple platforms to reach distinct audiences, in order to control who sees or does not see their posts \cite{litt_just_2016, marwick_i_2011, zhao_social_2016}. 

\section{Study Design}
\label{sec:methods}

To study overlapping membership in online communities, we conduct interviews with members of online communities hosted on Reddit, a social media platform for sharing, discussing, and rating news, media, and other content in user-created subcommunities called ``subreddits.'' Individual users can participate in any of Reddit's millions of subreddit communities by posting ``submissions'' that might include a link to a news article, a question for discussion, an image, or text written by the submitter. Each submission has a corresponding threaded comments section. Users can also vote submissions and comments up or down as a form of distributed moderation and can give awards to comments and posts \cite{lampe_slashdot_2004, burtch_how_2021}.

Subreddit communities are managed by teams of volunteer content moderators tasked with curtailing abusive behavior and keeping conversation on topic \citep{matias_civic_2019, seering_moderator_2019}. 
Subreddits exist covering an enormous range of topics \citep{fiesler_reddit_2018}, and
Reddit has been the site of much research on overlapping online communities \cite{datta_identifying_2017, tan_all_2015, tan_tracing_2018, hessel_science_2016, teblunthuis_identifying_2021}.
Because the cost of creating and joining new communities on Reddit is very low, subreddits often overlap in both topic and membership. Users frequently create spinoff subreddit communities from larger and more established groups \cite{tan_tracing_2018}.

\subsection{Participant Selection}

To understand why people participate in overlapping communities, we set out to interview people who are active in highly related subreddits. Additional inclusion criteria were that users were adults (i.e., above the age of majority in their country) and able to participate in an interview in English.  

Our participant selection process began by first choosing clusters of highly related groups. To do so, we built a web-based data visualization of a clustering algorithm derived from user overlap to identify groups of interest-based subreddits having similar users.
To generate the visualization, we conducted a computational analysis of the Pushshift Reddit dump containing a nearly complete collection of Reddit comments made before April 2020 \cite{baumgartner_pushshift_2020}. We selected the top 10,000 subreddits based on the number of comments in this data and excluded subreddits where a majority of submissions were flagged as not safe for work. Next, following an approach described in prior work \cite{datta_identifying_2017}, we constructed the measure of user similarity by taking the cosine similarities of TF-IDF vectors. Using this similarity measure, we ran affinity propagation clustering \cite{frey_clustering_2007} to group subreddits having overlapping users.  We then built an HTML visualization of these clusters based on t-distributed stochastic neighbor embedding (t-SNE). We have included the visualization in our online supplement. 

Although some aspects of our manual cluster selection process using this visualization were necessarily arbitrary, we tried to select clusters that were interest driven, involved primarily English language discussion, and were focused on content about which all members of the research team would be comfortable speaking. As a result, we did not select any clusters that were focused on sex or pornography, fringe or extreme politics, content specific to geographic regions, or topics that our group could not understand. 

We sought out clusters that we hoped would result in individuals from a diverse range of ages, genders, and life experiences. Although we did not collect demographic information from our interviewees, our interviewees' presentation and descriptions of themselves suggested that these efforts were not entirely successful. 
Our pool of interviewees included young and middle-aged people; people of color; people from the United States, Canada, and Europe; people who did not speak English as a first language; and people who were not men. 
That said, men were very likely overrepresented in our pool of interviewees, perhaps even in relation to the disproportionate participation of men on Reddit \citep{amaya_new_2021}. 

The clusters we selected each include 3--10 subreddits on the following topics: rock climbing, streetwear fashion, roller coasters, vintage audio, podcasting, painting, drag culture and performance, indie music, and dating for middle-aged adults. Information about each subreddit and cluster can be found in Table \ref{tab:subs_clusters_stats}.

\begin{small}
\begin{table}[t]
    \centering
    \begin{tabular}{cccc}
\hline
  \textbf{Subreddit}    & \textbf{Cluster}  & \textbf{Subscribers} & \textbf{Created} \\
  \hline
\rowcolor{lavenderblue} 
r/bouldering          & Climbing          & 194,814              & 2009-10-28       \\ 
\rowcolor{lavenderblue} 
r/climbharder         & Climbing          & 117,288              & 2010-10-19       \\ 
\rowcolor{lavenderblue} 
r/climbing            & Climbing          & 935,621              & 2008-07-17       \\ 
\rowcolor{lavenderblue} 
r/climbingcirclejerk  & Climbing          & 45,032               & 2011-08-18       \\ 
r/Drag                & Drag              & 44,724               & 2011-01-15       \\ 
r/Dragula             & Drag              & 27,510               & 2016-11-03       \\ 
r/rupaulsdragrace     & Drag              & 440,329              & 2011-11-15       \\ 
r/RPDR\_UK            & Drag              & 31,867               & 2019-02-07       \\ 
r/SpoiledDragRace     & Drag              & 69,027               & 2018-02-16       \\ 
r/MsPaintsArtRace     & Drag              & 61,292               & 2017-04-17       \\ 
\rowcolor{lavenderblue} 
r/MGMT                & Indie Music       & 17,744               & 2010-02-25       \\ 
\rowcolor{lavenderblue} 
r/tameimpala          & Indie Music       & 94,248               & 2011-10-30       \\ 
\rowcolor{lavenderblue} 
r/kgatlw              & Indie Music       & 59,191               & 2015-07-01       \\ 
\rowcolor{lavenderblue} 
r/Indieheads          & Indie Music       & 1,932,698            & 2013-12-24       \\ 
r/datingoverthirty    & Middle Age Dating & 436,480              & 2014-11-04       \\ 
r/DatingAfterThirty   & Middle Age Dating & 11,550               & 2018-03-09       \\ 
r/datingoverforty     & Middle Age Dating & 52,522               & 2018-12-15       \\ 
r/relationshipsover35 & Middle Age Dating & 14,916               & 2018-02-06       \\ 
\rowcolor{lavenderblue} 
r/OilPainting         & Painting          & 186,716              & 2011-09-22       \\ 
\rowcolor{lavenderblue} 
r/Painting            & Painting          & 280,865              & 2008-06-13       \\ 
\rowcolor{lavenderblue} 
r/PourPainting        & Painting          & 178,800              & 2017-07-28       \\ 
\rowcolor{lavenderblue} 
r/Watercolor          & Painting          & 269,882              & 2012-01-15       \\ 
\rowcolor{lavenderblue} 
r/HappyTrees          & Painting          & 53,362               & 2011-02-07       \\ 
r/podcasts            & Podcasting        & 1,995,693            & 2008-01-25       \\ 
r/podcast             & Podcasting        & 60,497               & 2009-01-02       \\ 
r/podcasting          & Podcasting        & 73,010               & 2010-09-17       \\ 
r/audiodrama          & Podcasting        & 129,102              & 2010-11-30       \\ 
r/ska                 & Podcasting        & 34,397               & 2008-03-12       \\ 
\rowcolor{lavenderblue} 
r/guessthecoaster     & Rollercoasters    & 5,094                & 2017-06-30       \\ 
\rowcolor{lavenderblue} 
r/rollercoasterjerk   & Rollercoasters    & 12,378               & 2016-07-14       \\ 
\rowcolor{lavenderblue} 
r/rollercoasters      & Rollercoasters    & 66,652               & 2010-07-31       \\ 
\rowcolor{lavenderblue} 
r/rct                 & Rollercoasters    & 55,275               & 2010-08-04       \\ 
\rowcolor{lavenderblue} 
r/themeparkitect      & Rollercoasters    & 13,536               & 2014-06-16       \\ 
r/streetwear          & Streetwear        & 2,678,745            & 2011-04-30       \\ 
r/supremeclothing     & Streetwear        & 154,797              & 2012-04-04       \\ 
r/womensstreetwear    & Streetwear        & 421,279              & 2016-04-25       \\ 
r/bapeheads           & Streetwear        & 19,672               & 2013-08-12       \\ 
r/malefashion         & Streetwear        & 207,843              & 2011-04-02       \\ 
r/sadboys             & Streetwear        & 74,932               & 2013-06-30       \\ 
r/techwearclothing    & Streetwear        & 94,675               & 2017-03-01       \\ 
r/Vans                & Streetwear        & 51,997               & 2011-07-01       \\ 
\rowcolor{lavenderblue} 
r/cassetteculture     & Vintage Audio     & 45,615               & 2011-05-25       \\ 
\rowcolor{lavenderblue} 
r/typewriters         & Vintage Audio     & 20,037               & 2010-10-25       \\ 
\rowcolor{lavenderblue} 
r/vintageaudio        & Vintage Audio     & 59,202               & 2011-09-18       \\ 
\hline
    \end{tabular}
    \caption{Clusters of subreddits from which we recruited participants, subscriber counts at the time of the study, and the creation date of each subreddit.}
    \label{tab:subs_clusters_stats}
 \end{table}
\end{small}

Using the Pushshift Reddit dataset, we identified candidate participants who were among the top 80\% most frequent commenters within each cluster, who participated in multiple subreddits in the cluster, and who were active in the cluster during a period of at least 1 calendar year. 
We began recruiting a random sample of 50 candidates matching these criteria within each cluster by sending direct messages through Reddit. Interested potential recruits filled out a short online survey confirming that they were adults and able to participate in English language interviews. The survey also asked participants about their participation and familiarity with each of the subreddits in each cluster to verify that they were knowledgeable. 
At the beginning of each interview, we asked if there were any other subreddits related to those identified by the clustering algorithm. As a result, our conversations were not limited to the subreddits listed in Table \ref{tab:subs_clusters_stats}.

We began by recruiting participants from the first three clusters listed in Table \ref{tab:subs_clusters_stats}. We found ourselves reaching saturation within these clusters quickly. We also found that different clusters were surfacing quite different data. In response, we added additional clusters and recruited at least two participants from each until we reached global saturation.  In some clusters, we did not reach saturation in two interviews. In these cases, we sent additional invitations and conducted additional interviews. 
In total, 20 participants were successfully recruited and interviewed by five members of the research team before we reached global saturation and ceased data collection. The characteristics of our interviewees are presented in Table \ref{table:participants}.

All of our interviews were semistructured. Although we drew from a long series of open-ended questions about participation in different subreddits and the relationships between communities, we chose our questions based on what our subjects wanted to talk about. A copy of our interview protocol is included in our supplementary material.
Interviews were 49 min long on average but varied substantially in length. We suggested conducting interviews over Zoom but offered the participants their choice of communication channel. As a result, we conducted two interviews over the phone, one using Discord chat, and the rest over Zoom.
Interviews were transcribed automatically using Zoom's built-in transcription and the otter.ai service and were then manually corrected by the authors. After each interview, participants were compensated with a digital gift card for \$20 USD through the Tango Card reward service.\footnote{\url{https://www.tangocard.com/}}

\subsection{Qualitative Data Analysis}

Our analysis followed \citepos{charmaz_constructing_2015} approach to grounded theory as closely as possible. We conducted coding and data collection in parallel. We generated over 950 codes, which we then grouped in an iterative axial coding process that generated 18 thematic memos.  As we completed collecting data, we refined our codes and combined themes to identify answers to our following orienting research questions: Why are there so many similar online communities? And why not more? Although primarily inductive, our analysis was influenced by sensitizing concepts from prior work including our knowledge of scholarship on overlapping online communities described in §\ref{sec:overlapping} and the reasons that people participate in online communities summarized in §\ref{sec:reasons}.
In analyzing our data, we noted that interviewees described their participation in multiple different subreddits and their preference for particular subreddits in terms of the inability of one community (often the ``main'' or ``largest'' community) to provide the desired benefits. This observation forms the basis of the grounded theory around which we organize our findings. 
All parts of this process were completed by the full team.

\begin{table}[t]
\small
    \centering
    \begin{tabular}{ccc}
\hline
\textbf{Participant ID} & \textbf{Cluster}     & \textbf{Interview Length (min)} \\ \hline
\rowcolor{lavenderblue} 
C1                      & Climbing             & 56                                  \\ 
\rowcolor{lavenderblue} 
C2                      & Climbing             & 51                                  \\ 
\rowcolor{lavenderblue} 
C3                      & Climbing             & 41                                  \\ 
D1                      & Drag                 & 51                                  \\ 
D2                      & Drag                 & 67                                  \\ 
\rowcolor{lavenderblue}
I1                      & Indie Music          & 71                                  \\ 
\rowcolor{lavenderblue}
I2                      & Indie Music          & 43                                  \\ 
O1                      & Podcasting           & 30                                  \\ 
O2                      & Podcasting           & 44                                  \\ 
\rowcolor{lavenderblue} 
P1                      & Painting             &  58                                   \\ 
\rowcolor{lavenderblue} 
P2                      & Painting             & 35                                  \\ 
\rowcolor{lavenderblue} 
P3                      & Painting             & 40                                  \\ 
\rowcolor{lavenderblue} 
P4                      & Painting             & 35                                  \\ 
R1                      & Rollercoasters       & 24                                  \\ 
R2                      & Rollercoasters       & 43                                  \\ 
\rowcolor{lavenderblue} 
S1                      & Streetwear           & 79                                  \\ 
\rowcolor{lavenderblue} 
S2                      & Streetwear           & 55                                  \\ 
T1                      & Dating in Middle Age & 63                                  \\ 
T2                      & Dating in Middle Age & 53                                  \\ 
\rowcolor{lavenderblue} 
V1                      & Vintage Audio        & 34                                  \\ 
\rowcolor{lavenderblue} 
V2                      & Vintage Audio        & 56                                  \\ \hline
    \end{tabular}
\caption{List of anonymized participant IDs, the cluster from which we recruited them, and the length of their interview.}
\label{table:participants}
\end{table}

\subsection{Ethical Considerations}
Our study design was reviewed by the Institutional Review Board (IRB) at the University of Washington and was determined to be exempt. As part of the design of this study, we took several steps to protect the privacy of our research participants.  Participants were fully briefed about the design of the study before being interviewed and were given documents concerning the study and contact information for our IRB. Explicit consent was obtained from every participant.

Because this project involved collaboration with a relatively large team, we used the Keybase end-to-end encryption service for all discussion and data sharing.
Finally, participants were anonymized so that no direct identifiers were recorded in the process of data collection, and only anonymized pseudonyms (e.g., C1, P2, and V2, as show in Table \ref{table:participants}) are published in this paper. We made several minor edits to quotes to obscure potentially identifying details.

\section{Findings}

Why do people participate in multiple online communities around the same topic? The answer that emerged from our grounded theory is that no one community can provide all the benefits that users want. At a high level, we find that people have multiple and diverse motivations for participation in online communities. In §\ref{sec:benefits}, we describe the types of benefits they seek organized into three categories: (a) engaging with specific types of content, (b) homophilous socialization, and (c) sharing content contributions with as large an audience as possible. 
In §4.2, we use data from our interviews to describe the tensions between these benefits.
We also investigate how our interviewees understood competition and mutualism---key concepts from ecological studies in social computing---between overlapping communities. Our interviewees overwhelmingly found mutualism to be more consistent with their understandings of overlapping online communities than competition.
Our contribution comes in the form of a theoretical framework, grounded in our data, that describes how the full benefits of participating in communities can only be satisfied by portfolios of communities.

\subsection{Benefits Users Seek from Communities}
\label{sec:benefits}

\subsubsection{Specific kinds of content}
\label{sec:content}

Content on Reddit is organized into subreddits that define their own topical boundaries. These boundaries may be broad (e.g., news) or narrow (e.g., types of painting media). Moreover, subreddits that prohibit types of content or behavior generate niches for subreddits with different rules. Despite such forms of specialization, multiple communities often welcome the same content and encourage users to ``cross-post'' material.

A subreddit's topic---what it is about and what content should be posted---is often signified by its name. A climbing enthusiastic explains:

\blockquote[C1]{I think the name itself [\texttt{r/climbharder}], kind of specifically points out that: this is not for people who climb hard. It's for people who climb and want to climb hard\textit{er}. 
}

\noindent C1 describes how the purpose of a subreddit is tied to its name by emphasizing the adjectival suffix ``-er'' as indicative of the fact that the subreddit is not about achieving elite performance but about improving.

Similarly, a participant in subreddits about drag performance invokes Marshall McLuhan to describe how they know what content to post and where to post them:

\blockquote[D1]{ Let’s say you were a drag artist and you wanted to show off something that you just created. You would have to go select which community you wanted to show it off in. And I guess among those, [\texttt{r/Drag}] would be the one to do that in. But if you’re---if you’re wanting to show off a piece of artwork or something that you made of a queen from Rupaul's drag race---and the best place to show that off would be to go to [\texttt{r/rupaulsdragrace}] and post it there. So it’s [a] `the medium is the message' kind of thing. \ldots You know where would get the most views [and] where would be the best place to post your content.}

\noindent Like D1, our informants had deep knowledge of what kinds of specific content would be appropriate for each subreddit in their cluster.

Specialization also occurred as a form of regulatory arbitrage when one community had formal or informal rules about the kind of content that was allowed. In these cases, we would often hear about an adjacent community where breaking the rules is accepted, perhaps even the raison d'être. For example, \texttt{r/rupaulsdragrace} prohibits spoilers and information about the outcomes of a reality TV show. \texttt{r/spoileddragrace} is a community about the same show that allows spoilers.

This pattern is so widespread on Reddit that it is often signaled in subreddit naming conventions \citep{hessel_science_2016}. The ``meta'' prefix signals meta-discussions, often drama-centered, about another subreddit. The ``jerk'' suffix signals a space for memes, mockery, silliness, or other content unaccepted in the ``main'' subreddit. Both are commonly understood and were discussed at length by our interviewees. 
For example, among the Rollercoasters subreddits, R1 described the ``jerk'' subreddit as a ``joke subreddit'' where members of the main rollercoasters subreddit could make fun of themselves:

\blockquote[R1]{I would definitely say \texttt{r/rollercoasters} and \texttt{r/rollercoasterjerk} are really deeply intertwined. It's usually all the same members and stuff because of the fact that the coaster `jerk' is just meant to make fun of the main subreddit. It's just a joke subreddit.}

\noindent ``Jerk'' subreddits were a common source of discussion among our participants. 

Among the Climbing subreddits, the ``main'' subreddit about rock climbing (\texttt{r/climbing}) is welcoming to newcomers. C1 explained that members upvote posts by newcomers``to encourage more entrance into the sport.'' However, newcomer posts are often repetitive pictures of people climbing in gyms or videos of famous climbers. This annoys some experienced climbers. The ``jerk'' subreddit provides a backstage space where making fun of newcomers is permitted.

In addition to being divided by rules, interrelated subreddits can be structured as a ladder of ``conceptual rungs'' where one finds larger communities as one ascends the ladder. A participant in the subreddits on art and painting described this phenomenon as

\blockquote[P2]{
You go up through these conceptual rungs.
\ldots\ 
When you go up from, say, \texttt{r/OilPainting}---like \texttt{r/HappyTrees} to \texttt{r/OilPainting}---it’s a much bigger community. And then from \texttt{r/OilPainting} to \texttt{Painting}, which is even bigger.
}

\noindent P2 explained that smaller subreddits such as \texttt{/r/HappyTrees} support learners and are generally more welcoming places. Although the quotation above suggests that the size of communities increases as one moves up conceptual rungs, the relationship between topical scope and size was more complicated. In some topical areas, subreddits with relatively specific topics have the largest and most active communities. For example, \texttt{/r/rupaulsdragrace} is the most active drag subreddit by a large margin, even though it focuses on a reality TV series that is part of the broader drag community covered by \texttt{r/drag}.

Although many specialized subreddits exist, people who want to share their work, ask a question, or have a specific discussion may not know the best place to post. Cross-posting---i.e., when someone posts the same content, questions, or messages in multiple communities---is widespread on Reddit.
Cross-posting has sometimes been viewed negatively as a form of attention grabbing (i.e., ``karma whoring'') \citep{poor_mechanisms_2005}.
More often, however, we heard that cross-posting was acceptable and even encouraged to establish complementary conversations or find different audiences. 

Multiple interviewees from the Climbing cluster, including C1, described how, when people ask for training advice in \texttt{r/climbing}, the largest subreddit about rock climbing, they will be advised to cross-post to \texttt{r/climbharder}:

\blockquote[C1]{Somebody will post asking for advice in \texttt{r/climbing} and oftentimes, somebody will comment and be like, `Hey, you know? You’re welcome to ask this here, but you might get more and better responses at \texttt{r/climbharder}.'}

\noindent C1 explained that even though conversations about training often start in the main subreddit, they are not likely to gain traction because not everybody in the main community is interested in the more intensive aspects of climbing.  

In sum, the ecosystem of subreddits about similar topics provides more opportunities for people to find specific desired discussions. People receive positive feedback and engagement when they post content that fits a subreddit's specific topic. That said, the subreddit where a particular piece of content will be best received is often not clear to the person posting it. Cross-posting provides multiple chances to start a desired discussion. 

\subsubsection{Homophily}

Online communities have long been recognized as a way to ``find my people'' by bringing together users who share things as diverse as a psychiatric diagnosis, enthusiasm for a hobby, or membership in a subculture or identity group. A member of the Middle Age Dating cluster of subreddits explains:

\blockquote[T2]{[When I joined the ADHD Reddit sites], I feel like I found my people after all these years.
\ldots\
If you don't have ADHD, and don't wonder what's going on other people's brains all the time, I think you just think that everybody thinks like you. And they don't. They don't. So if you're 30 and you're having a problem, you really just want to talk to other 30 somethings.
}

\noindent T2's description of having ``found my people'' and talking to other people like themselves invokes the idea of homophily: the desire to connect to others similar to oneself. 
Analytically distinct from finding personalized information in narrowly focused subreddits, homophily was frequently cited as an end in itself by our interviewees.
Our interviewees sought to connect with ``like-minded'' people having similar interests, demographics, identities, tastes, and status.

Even though the identities of others in subreddit communities are largely invisible, participants can easily imagine the demography of the subreddit. A participant in the Drag cluster of subreddits described \texttt{r/Dragula}, a community of fans of a TV show featuring horror-infused drag styles, as follows:

\blockquote[D2]{ 
I think it would be a mostly LGBTQ audience. And not many straights. But if there are straights, they would be really open minded or edgy. Or, I don’t know \ldots\  associated with that `dark' aesthetic.}

\noindent D2's thoughts on \texttt{r/Dragula} convey a clear sense of the audience of the subreddit. Of course, the pseudonymous nature of Reddit obscures age, race, gender, and ethnicity. That said, Reddit users draw on stereotypes about fanbases and cues such as mentions of schools, selfie posts, linguistic markers, and cultural references to build clear models of the types of people in a subreddit. In further unpacking these dimensions, D2 contrasts \texttt{r/Dragula} with the more mainstream subreddits about the show \textit{Rupaul's Drag Race}: 

\blockquote[D2]{ [As for subreddits about] the drag race (\texttt{r/rupaulsdragrace}), Drag Race UK (\texttt{r/RPDR\_UK}), and the spoiled drag race (\texttt{r/SpoiledDragRace}). \ldots\ 
Most of [the participants in these other groups] don’t do drag. Most of them are, I think, white gay men, or straight women who see drag with a very narrow view of what drag is. Hegemonic? I don’t know if that’s the word, but they apply the same standards of beauty that are applied to women and men and artists and performers to this art form. 
}

\noindent D2 conveyed both a strong sense of the demographics of different drag subreddits and a strong sense of identification with \texttt{r/Dragula}, which they described as less toxic, more inclusive, and more creative, in part because its membership has a greater concentration of LGBTQ and non-White people who are less interested in conforming to hegemonic beauty standards.  

Subreddits divide broad topical areas such as drag, art, and fashion into subgroups of people occupying strata of status hierarchies associated with identity, expertise, and class. For example, in the Climbing cluster of subreddits, rock climbing ability confers status and separates beginners from advanced athletes. We found that these two groups concentrate their participation in different subreddits. Across the clusters, we found that experts sought out fellow experts with whom to share knowledge, offer reflections, and give advice grounded in shared extensive experience.

Our Streetware interviewees reported that subreddits about fashion are split along lines that are associated with the price and status of the clothes being discussed:

\blockquote[S1]{The kind of person, the Platonic ideal poster or user of something like \texttt{r/streetwear}, is probably more open-minded, maybe, in terms of what they think is cool, what they think is worth wearing. Whereas, you know, \ldots\ the \texttt{r/malefashion} snob is a snob.}

\noindent Even though users of \texttt{r/streetwear} share and discuss men's fashion, \texttt{r/malefashion}, which focuses on higher-status and more expensive styles, looks down on their casual and youthful styles.
S1 is a member of the \texttt{r/streetwear} subreddit. 
Although their groups are ``chill'' and ``supportive,'' higher-status groups are ``snobby.'' It is clear that S1 feels unwelcome and out of place in the higher-status group.

Similarly, our interviewees described status hierarchies in Painting subreddits related to skill level and medium.
P4 described how they were invited to cross-post their work from \texttt{r/Watercolor} to \texttt{r/Artoilpainting}, a smaller subreddit that seems to have a complicated relationship with watercolor.  Although watercolor submissions are allowed, and, in this instance, encouraged, both the subreddit's name and the similarity between its visual tag for watercolor submissions with the downvote button suggest that oil is the preferred medium in this community.
In this way, the division of topical spaces into spheres of similar status and identity allows members to find groups that exclude both those who look down on them and those who they look down upon.

Although ``finding your people'' is satisfying in itself, it can also be a foundation for a wide range of other kinds of benefits. 
For example, a homophilous community leads to conversations that can promote trust. Trust has many benefits such as building confidence in the advice and information shared within a community. In some communities we studied, this trust enabled buying, selling, and trading of material goods.

V2, one of our interviewees from the Vintage Audio cluster, described a community of record collectors on Reddit that acted as a market for buying, selling, and trading records. They preferred this subreddit to other online markets such as Ebay because the community holds members accountable for honest transacting and because of the intrinsic reward that comes from sharing records with a fellow community member:

\blockquote[V2]{Because it's a group of people that are like-minded, \ldots\ your feet are kind of held to the fire a little bit more about actually being realistic with the condition [of the material you are selling]. Whereas, [when you buy] vinyl at the used record shop, sometimes you feel like someone's trying to pull one over on you \ldots\ I feel like because it is a community, sometimes you can get some kind of better deal \ldots\ \\ I found other people that share the hobby that I like. So I almost, definitely, feel like they’re friends in a little way. And so I want to, if I’m ever selling, I’m going out of my way to make sure that whatever I’m doing, everything I’m doing, is above board.
}

V2 was very enthusiastic about the ``marketplace wrapped in a community'' for vinyl records. According to V2, both buyers and sellers of records benefit from transacting within a community of like-minded hobbyists. Because the community holds sellers accountable, the community promotes honest representation of merchandise.  Being part of a like-minded community where members feel friendship with each other gives sellers a reason to be honest, and even to discount their wares, because they get ``some kind of better deal.''

In sum, our interviewees turned to specific subreddits to find people who share their interests, tastes, problems, and identities.
Our participants described subreddits in terms of demographics and identity groups as well as styles, subgenres, or categories related to social status such as wealth, expertise, and beauty standards. They used these categories to place themselves within the constellation of related subreddits they participated in.
Members of subreddits who are ``finding their people'' benefit each other by acting as communities as well as building trust and feelings of friendship. Over time, these feelings can provide further benefits such as the ability to more safely engage in buying and selling. 

\subsubsection{Finding the largest possible audience}

A third type of benefit derives from the number of members in a subreddit.
All our interviewees were keenly aware of the fact that a post reaching one of the top positions on a larger subreddit would receive the attention of a vast audience.  They described this attention as emotionally thrilling and otherwise beneficial. For artists and influencers, large audiences brought material rewards. For learners, a large audience's collective knowledge could bring hard-to-find answers and advice.

That said, our interviewees explained that larger subreddits do not necessarily provide a larger audience because posts in larger subreddits are more likely to be ignored or missed in the torrent of other content. Although posting in a smaller subreddit might increase the chances of finding an audience at all, subreddits that were too small were described as unattractive because they would not attract many posts or replies. Interviewees responded by choosing where to post strategically.

Although the competition for the top spots on the front page of large subreddits can be fierce, this competition can make recognition from a large subreddit extremely gratifying:

\blockquote[P2]{Likes are just kind of fake: fake social currency. But yeah, when you get a charge out of it, yeah, I love it. Most of the time, painting is a really busy sub. I mean, like, in any given hour, the new page is already replaced.
\ldots\ \\ 
If you can get something that gets a hold there and stays on the front page for a little while, [if] it gets up in even the top five, I've had a handful do that. That's kind of cool. 
}

\noindent P2 describes the thrill of reaching top positions in \texttt{r/painting} with posts of their paintings.  Even though they are dismissive of likes on Reddit, they desire the attention their work gets from the subreddit. It sends traffic to their websites, raises their artistic profile, and helps them sell their art. Although these material incentives are important, part of the thrill comes from knowing that a given subreddit is competitive. Smaller subreddits are simply unable to provide these benefits.

However, posting in a large subreddit also means the risk of being ignored:

\blockquote[S2]{I think there’s this weird bell curve where the community needs to be big enough where people want to post content. But it can’t get too big where people are drowning each other out for attention.}

\noindent S2 was among several of our interviewees who described an ideal ``middle ground'' for subreddit size. In general, we heard that people were less interested in posting content in very small subreddits that do not provide an audience. Thus, competition over the largest audiences drives people to smaller subreddits where they can reliably find an audience. I2 from our Indie Music cluster explained:

\blockquote[I2]{Usually \texttt{r/Indieheads} is the way to reach more people if you want to. Just like if you wanted to do even more, you’d probably do it on \texttt{r/music}. \ldots\  Say a small indie band decided to do an AMA they would probably want to do it on \texttt{r/Indieheads}. Because if they did it on \texttt{r/music}, it would get drowned out and nobody would see it because there’s so many posts. In \texttt{r/Indieheads} it would get a decent bit of attention, I think. In the band subreddit, it would probably get a lot of attention too. But \texttt{r/Indieheads} seems like the best middle ground for that kind of thing.
}

\noindent I2 explained that when the psych-rock band \textit{King Gizzard and the Wizard Lizard} wanted to engage with an audience on Reddit, they had a choice whether to post in the smaller ``band subreddit'' dedicated to them, the very large \texttt{r/music}, or the medium-sized \texttt{r/Indieheads}. Although posting in the band subreddit would have surely provided an audience, they chose \texttt{r/Indieheads}, which was large but where there was still little risk that their post would be drowned out.  

Our interviewees repeatedly described how finding an audience for one's content is a clear motivation for posting in larger subreddits.
However, we also heard that competition for attention in the largest subreddits leads people to try to find an audience in smaller subreddits. 
In the smallest subreddits, posting may not seem worthwhile at all.
This trade-off between finding a large audience and being ignored suggests that posting in subreddits of intermediate size can be the most reliable way to reach a sizable audience. 

 \subsection{Tensions Between the Benefits}
\label{sec:tradeoffs}

The findings in the previous sections imply a clear reason that so many overlapping subreddits exist. When one subreddit prohibits a certain type of content or conversation, an adjacent group can form that allows it. When an identity group is marginalized in one subreddit, members of that group may form a subreddit of their own. When getting attention in a large subreddit is too difficult, a smaller subreddit becomes attractive.
Using data from our interviewees, we describe each of the three possible tensions that exist between the three benefits: (1) subreddits where one finds a large audience are less able to provide specific types of content; (2) communities with large audiences are rarely able to provide a community of similar others; (3) some valuable types of discussion and information are found only in diverse groups of people.  As we discuss in §\ref{sec:discussion.trillemma}, taken together, these tensions form a ``trilemma''---i.e., a choice with three mutually incompatible options---between our interviewees' desires for specific content, homophily, and finding audiences. A single community might provide two of these benefits, but almost never all three. 

\subsubsection{Larger audiences create background noise}

In §\ref{sec:content}, we described how subreddits are structured according to distinctions between different types of content. Breaking topical areas into subreddits of varying levels of granularity makes finding specific content easier because doing so reduces the need to sift through unrelated material in a large and broad subreddit. Our interviewees often expressed that larger subreddits are simply not the best places for enthusiasts to have discussions:

\blockquote[C2]{I see this background noise problem building [in] \texttt{r/climbing}, the main climbing community, [which] has just become less and less and less interesting and less relevant as it’s gotten bigger. That’s not really a problem. Right? That’s probably has more to do with my interest level and how long I’ve been on it. And my experience level with climbing. I'm just a little bit more crusty about it, you know?
}

\noindent C2 describes losing interest in the primary subreddit about climbing as it grew because of the interviewee's specific interest in particular types of climbing content (i.e., material associated with being ``crusty'' or experienced). C2 recognizes that when \texttt{r/climbing} experienced growth, the larger volume of posts by newcomers to the sport created a ``background noise problem'' that made it difficult for established climbers to find discussions of interest.

Similarly, smaller subreddits can be incredibly valuable to those looking for highly specialized information.  Even though they may have very low levels of activity, they can provide a way to learn about rare forms of expertise. A participant in our Vintage Audio cluster explained how they might seek out advice on building a reel-to-reel audio setup: 

\blockquote[V2]{
If you're at [\texttt{r/ReelToReel}]. Everybody is hyper into them. Whereas there's probably overlap with somebody in  \texttt{r/vintageaudio} \ldots\ If I'm like trying to rebuild my reel-to-reel player, I want to talk to \ldots the most knowledgeable person particularly about building reel-to-reel \ldots
So I know that who I'm talking to is hyper specific to the knowledge I want. 
}

\noindent Invoking \texttt{r/ReelToReel}, V2 describes a highly niche subreddit about archaic audio tape equipment with only 3,200 subscribers and a handful of posts each day. V2 is simply not looking to find a large audience. Instead, they want access to the ``most knowledgeable person'' with specific expertise because access to this expertise makes it possible for them to consider doing their own reel-to-reel projects. 

Although the \texttt{r/ReelToReel} community overlaps with the larger and more general \texttt{r\Slash vintageaudio}, the latter does not provide the ability to connect with a small group of enthusiasts who are expert in an old-fashioned technology. 

Similarly, when someone wants a podcast recommendation tailored to their personal tastes, asking in a larger subreddit is not likely to prove as fruitful as it is within a smaller one. O2, a participant in the Podcasting cluster explained:

\blockquote[O2]{So I think for like \texttt{r/audiodrama}, I would probably write a longer post, and probably get a bit more into like, my personal tastes. Like I would comment about, `oh, I really love the acting in this one, is there anything similar?'
Open up a bit more about what I do and don’t like. Whereas I think in podcasts, it probably would be more  direct. I’d ask a specific question \ldots\ more to the point, more factual, probably just more almost transactional.}

\noindent Although the larger \texttt{r/podcasts} subreddit is a popular place to promote podcasts on Reddit, O2 explains that they prefer asking for recommendations in the smaller \texttt{r/audiodrama} where they find others willing to take their personal tastes into account. Our interviewees did not advance a ``smaller is better'' argument. O2 explains that they still engage in larger subreddits but use a more direct and transactional approach to information exchange when they do. Similarly, large art communities provide opportunities to find a large audience, but someone can find more substantive feedback to improve their skills, by posting in a smaller subreddit organized specifically for this purpose.

Interviewees described the most general interest-based subreddits such as \texttt{r/podcasts}, \texttt{r/painting}, and \texttt{r/climbing} as more accessible and welcoming to newcomers and as reaching a larger audience: all things they valued. They also described these larger groups as having a high volume of low-effort posts or comments. 
Our interviewees explained that although they play a useful role in an information ecosystem, the largest subreddits in a topical area are rarely the best places to look for information or advice.
They explained that small subreddits can effectively play host to content, information, and discussion that larger subreddits cannot.

\subsubsection{Homophily is more difficult in larger groups}

Because they have less background noise, smaller subreddits are more likely to provide better opportunities to connect with people who share one's distinctive interests, tastes, and identity.  Smaller subreddits are also better places to find a community because they provide opportunities to have repeated encounters with recognizable others, off-topic discussions, and personal interactions. P4 explained:

\blockquote[P4]{Obviously, I want as many people to see my stuff as possible, especially [since I am] trying to establish myself. But at the same time, I do want to build a relationship with any sort of community that I can.}

\noindent P4 explained that they participated in multiple communities because they have two goals as an artist. First, they want to find an audience for their artwork to establish their career. Second, they want to build a community with others who share their craft. They felt that they needed to turn to multiple subreddits to satisfy both needs.

Although larger subreddits provide a large potential audience, smaller subreddits were described as being friendlier. Another interviewee from our Painting cluster explains that this is because of how people act differently in large and small subreddits:

\blockquote[P3]{
I live in the middle of nowhere. And every so often, before the pandemic, I would visit [the large city several hours away]. Now I found there were very polite people, both in [the city] and in [my rural area]. But the tone by which people carried themselves changes in their environment. That's kind of one of the big changing factors. So, in the city, people are in a rush. They're about their business. We don't really have time to chat. 
\ldots\ 
The big subreddits might seem unfriendly [but] it’s not that so much. Individual members are impolite or unfriendly. But it’s almost as though people carry themselves differently when we’re in different subreddits.}

\noindent In their extended metaphor, P3 explained that large subreddits are like big cities full of busy people who do not ``have time to chat.''  Evocatively, they described people as behaving differently in large and small subreddits. The very same people who are rude in large subreddits might be friendly in smaller subreddits where people have repeated encounters with one another and have a stronger sense of knowing each other.
In another quote from the same cluster, P2 described how the small subreddit for Bob Ross-inspired painters, \texttt{r/HappyTrees}, stands out from the larger art subreddits because people know one another and it does not feel anonymous. The tight-knit nature of this community contributes to its utility as a source for feedback.

\subsubsection{Tension between finding specific content and homophily}

A third tension described by our interviewees is between the desire to connect with similar others and the desire for forms of discussion, content, and feedback that can only be found in diverse groups inclusive of dissimilar others.
Our interviewees described a range of situations when they sought out dissimilar others. For example, they described beginners seeking to learn from experts and outsiders seeking to learn about other cultures. They also described how subreddits instituted rules to limit or organize content that also interfered with unstructured and off-topic discussions that helped with community building.

For example, although multiple subreddits with overlapping users discuss the same episodes of the TV series \textit{Rupaul's Drag Race}, they have different understandings of events in the show depending on their national identities. D1 explained:

\blockquote[D1]{
The discussions played out differently on different subreddits. In the Drag Race UK sub there’s a lot more understanding about [a British drag queen] in particular, about where they come from \ldots\ In America we don’t understand how that person is from Worcestershire. }

\noindent D1 explained that the cultural background of one of the drag queens was a subject of discussion in \texttt{r/RPDR\_UK}, the UK drag race subreddit, while the main subreddit, \texttt{r/rupaulsdragrace}, was ``dominated by the American viewpoint.'' 

Our interviewees described a number of subreddits focused on discussing broad topics from a specific national or regional culture context. These cultural communities within a topical area provide a homophilous space for sharing distinctive cultural knowledge and sensibilities.
The wrinkle is that even for our American interviewee D1, the \texttt{r/RPDR\_UK} subreddit provided an opportunity to enhance their own experience and appreciation of the show by observing and learning from members of another culture.
In examples such as these, our interviewees explained that communities where like-minded people can share their distinctive appreciation could provide a source of knowledge for outsiders.

Similarly, Painting participant P2 explained that a group that has a mixture of experts and beginners provides a better learning environment than does a group of beginners alone:

\blockquote[P2]{If you can find a small group, with a small core of people who are particularly skilled, they sort of energize the group as a whole. \texttt{r/HappyTrees}, even though it's kind of a beginner subreddit, there's some people that posts there that are like, you know, Bob Ross instructors, or they've been doing this for years. And they've mastered that sort of \ldots\ ``happy trees'' thing.
}

\noindent P2 explains that part of what makes \texttt{r/HappyTrees} great is that it connects learners to experts. A homogenous subreddit of only beginners or experts would not provide the same opportunities.

To stay focused on specific types of content, subreddit moderators will frequently employ strict rules and heavy-handed moderation. 
Our respondents explained that smaller subreddits can get by with fewer rules and lighter moderation because they have fewer behavior problems and are less attractive to toxic outsiders. They are also more able to self-police using Reddit's voting system and through direct interpersonal sanctions such as admonition. In the words of one of the Vintage Audio participants,

\blockquote[V2]{In Reddit, the more users you get, the more strict the rules, and the more strict the moderation. Just to prevent problems.
}

\noindent V2 continued and explained that when a subreddit is small enough that you can ``wrap your hands around'' it and is built around a ``like-minded'' group, it can develop and enforce shared behavioral norms that substitute for formal rules and rigid enforcement regimes. V2 explained that the processes of creating spaces for specific types of information got in the way of building community. 

Similarly, one of our interviewees described \texttt{r/Indieheads}'s rules limiting how often one can post, requiring specific titles and tags, and prohibiting types of user-generated content. Although these rules help maintain a high-quality feed, they also prevent the sharing of more personal and relatable forms of content such as amateur performances and chit-chat. As a result, subreddits that make rules to ensure that posts are on-topic frequently have adjacent ``-jerk'' subreddits that provide an outlet for jokes and memes and act as places where off-topic discussions can thrive.

\subsection{Interviewee's Understandings of Competition and Mutualism}
\label{sec:results.competition}

Except for a small qualitative subpart of a single paper \citep{zhu_selecting_2014}, prior ecological studies in social computing have relied on concepts such as competition and mutualism but have provided limited evidence that such concepts are salient to participants. 
As part of our interviews, we asked our interviewees if they perceived relationships between the communities they participated in to be competitive or mutualistic.  In some cases, interviewees imagined hypothetical scenarios where competition might emerge from the perspective of subreddit moderators. For example, a participant in Climbing said:

\blockquote[C1]{I guess if you put your Reddit [moderator status] on your resume or something, and you want to be a moderator of a larger community, you could try to get users from other communities. But I haven’t seen or experienced competition.}

\noindent Although we asked nearly every interviewee about competition, only one interviewee (S2) described an actual instance of conflict or direct competition.
In nearly every other interview, our subjects found our suggestion that subreddits might be in competition to be surprising and strange.

However, the idea that communities are complementary and mutualistic was much more intuitive. One Vintage Audio participant explained the relationship between subreddits:

\blockquote[V2]{Yeah, the overlapping. \ldots\ They each have their own niche. \ldots\ They get big enough to have super critical mass of people. Then they'll have a reason to exist. And then they'll sort of fit into the ecosystem of different communities.}

\noindent Consonant with this description of subreddits in unproblematic coexistence, our interviewees repeatedly suggested that there were not meaningful structural or technical limitations on the number of subreddits a user can join and this reduced the possibility of competition, if it did not eliminate it altogether. 

 \section{Discussion}

\label{sec:discussion.trillemma}
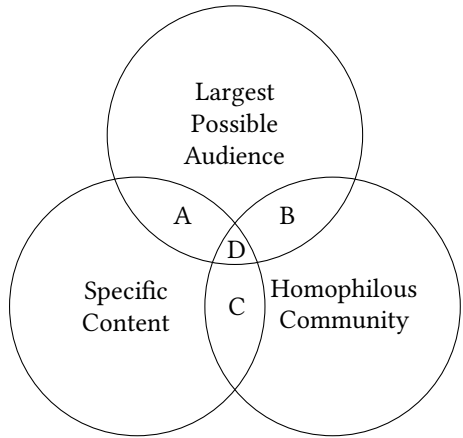
\begin{figure}
\def\firstcircle{(0,0) circle (1.7cm)}
\def\secondcircle{(60:2.6cm) circle (1.7cm)}
\def\thirdcircle{(0:2.6cm) circle (1.7cm)}

\definecolor{myyellow}{HTML}{fae772}
\definecolor{mygreen}{HTML}{4ac26c}
\definecolor{mypurple}{HTML}{31668c}
\begin{tikzpicture}
    \begin{scope}[shift={(3cm,-5cm)}, fill opacity=1, text width=2cm, text centered]

\draw \firstcircle node [xshift=-1ex] {Specific Content};
        \draw \secondcircle node [yshift=1ex] {Largest Possible Audience};
        \draw \thirdcircle node [xshift=1ex] {Homophilous \\ Community};
        \node (A) at (0.6,1.2) {A};
        \node (A) at (2,1.2) {B};
        \node (A) at (1.325,0.75) {D};
        \node (A) at (1.325,0) {C};

\end{scope}
\end{tikzpicture}

    \caption{Venn diagram illustrating the specificity-homophily-audience ``trilemma.''}
    \label{fig:trilemma}
\end{figure}

The tensions between the benefits that our interviewees sought can be thought of as forming a ``trilemma'' between finding specific content, homophily, and finding as large an audience as possible. This three-way dilemma captures the fact that the more a subreddit succeeds in providing any one of these benefits, the less able it will be able to provide the others. A portfolio of overlapping communities solves this problem by providing all three types of benefits.

Figure \ref{fig:trilemma} visualizes the theorized trilemma. Each of the benefits described in §\ref{sec:benefits} is reflected in large circles. Each of the tensions described in §\ref{sec:tradeoffs} is reflected in the overlapping areas in the figure.
Area A contains communities that provide the largest possible audience and specific content but are unlikely to provide homophily to community members. Subreddits that provide large audiences face ``the background noise problem'' as a large volume of submissions makes it difficult for people to find the specific content they care about.
Area B contains communities that offer both large audiences and homophily but that will struggle to provide specific content. For example, an American interested in learning about international drag culture finds the need to search beyond \texttt{r/rupaulsdragrace}.
Area C contains communities that provide specialized content and a homophilous community but that may not attract large audiences.
Although not everyone who desires a specific type of content may be similar to those who produce the content, smaller subreddits can often provide both desired content and opportunities to socialize with similar others. 
However, as the size of the audience increases, subreddits encounter the background noise problem and acquire a ``big city'' air of unfriendliness.

\subsection{Connections to Prior Research} 

\subsubsection{Finding specific content}

Our findings are consonant with prior work that the primary benefits provided by online communities stem from their power to connect people to novel and hard-to-find sources of information \cite{benkler_wealth_2006, campbell_thousands_2016, von_hippel_free_2016,fiesler_growing_2017}. 
Our study adds to this work and complements recent findings of \citet{hwang_why_2021} by describing how nested and overlapping online communities are useful for information seeking and managing one's information exposure.  Individuals often desire multiple types of content within a general subject area, such as spoiled and spoiler-free discussions.  
Even when a relatively obscure community such as \texttt{r/vintageaudio} exists, an even more specialized community such as \texttt{r/ReelToReel} may provide access to an even more specialized set of experts.

\subsubsection{Finding homophilous community}
Prior work has recognized the importance of homophily in motivating and structuring participation in online communities \cite{chang_specialization_2014, cunha_are_2019, grevet_managing_2014}. 
Contributing to this line of research, we identified a number of types of homophily that drive individuals' decisions to participate. These included hobbies, expertise, age, national culture, identity, and status. Homophily was in tension with the need for specific content in that differences among many of these dimensions were valuable for finding information.

Our results suggest that participants in online communities face trade-offs between homophily and information novelty. These may be similar in structure to the trade-offs between short and long ties observed in contexts such as work groups \cite{ruef_structure_2003} and social networks \cite{grevet_managing_2014, granovetter_strength_1973}. One advantage of joining a group of overlapping online communities is that it can help find information that would be unavailable in homophilous groups. 

\subsubsection{Finding the largest possible audience}

Much social computing research points to the benefits of large audiences and large communities \citep{kraut_building_2012}. Our work adds more evidence to back up those claims. More important, perhaps, are recent counterclaims about the benefits of smallness. \citet{hwang_why_2021} presents an interview study with members of small Reddit communities. Although our results about the tensions between large audience size and other benefits are fully in line with Hwang and Foote's findings, our starting assumptions and ultimate takeaways are quite different. \citeauthor{hwang_why_2021} seek to understand why people participate in persistently small communities and conclude that smallness offers a range of benefits. Our results suggest that individuals seek out benefits that happen to be incompatible with largeness and participate in portfolios of communities that, because of the trilemma we described, will almost certainly include small ones. Although we believe that \citepos{hwang_why_2021} emphasis on smallness might draw focus to a side effect instead of the cause, we believe that the findings in our two papers are largely complementary.

Although users may desire large audiences, large online communities often require additional structure to maintain order\cite{kiene_surviving_2016, gillespie_custodians_2018, kiene_technological_2019}. \citet{kiene_surviving_2016} describes how a massive influx of newcomers presents difficulties that can be managed by appointing additional moderators, increasing norm enforcement, and limiting the frequency of posts. \citet{lin_better_2017} find that such interventions help subreddits maintain comment quality and stay on topic during massive influxes of growth. Our sense is that these changes ensure the availability of specific content, in part, because of the growth-limiting effects of rules and enforcement \cite{halfaker_rise_2013, teblunthuis_revisiting_2018}. We see this as yet more evidence in favor of our theory.

\subsection{Implications for Ecological Studies in Social Computing}

The quote by V2 in §\ref{sec:results.competition} can be read as a kind of summary of resource partitioning theory (RPT), a strand of ecological research in organizational science that focuses on explaining specialization \cite{carroll_concentration_1985}. Although RPT has not been deeply examined in prior social computing work, our findings suggest that it may be able to explain the widespread occurrence of overlapping communities. RPT proposes that the reason that small specialized organizations coexist with large generalist organizations is that generalists are constrained in their ability to meet distinctive needs in niche markets \cite{carroll_why_2000, swaminathan_resource_2001}. In V2's terms, the ``ecosystem of different communities'' is constructed by a process in which those that ``have a reason to exist'' and are specialized to ``have their own niche'' will achieve ``critical mass.''

Our grounded theory suggests that the trade-offs in the capacity of an online community to provide different types of benefits that people seek from online communities give rise to new niches.
On the basis of our findings and our understanding of RPT, we hypothesize the following process to describe how systems of overlapping communities develop:

When a new topical area grows, the bulk of activity will happen in a generalist community. New members joining that community may seek and find the perceived benefits described in §\ref{sec:benefits} (i.e., specific kinds of content, homophily, and the largest possible audience). 
If a topic area, such as art, is sufficiently general, initial membership growth occurs as the community attracts new and existing users interested in both general and more specific types of content. 

As growth continues, membership in the generalist community becomes heterogeneous with lower levels of homophily (e.g., amateur and professional artists) and more specific interests (e.g., painters and photographers) and types of engagement desired (e.g., attention from an audience or critique). At this point, the trade-offs we discuss in §4.2 related to size become relevant. Finding information related to a specialized subtopic and homophilous socializing grows difficult.

If, as with Reddit, creating new communities is low cost, a community specialized in a subtopic can emerge.
This specialized community will likely not attract as large an audience as the generalist community. However, those most interested in the specific subtopic will join it to escape what our interviewees describe as ``background noise'' in the larger generalist community.
Similarly, those seeking personal interaction or social bonding with other community members will be more likely to find them in the specialized community.
A similar process occurs in the formation of spaces having different rules or purposes (such as ``jerk'' spaces). 
The cycle will then begin anew as subreddits repartition a subtopic such as \texttt{r/painting} into subspecialists such as \texttt{r/oilpainting} and \texttt{r/watercolor}.
Although some of our interviewees described parts of this process, the model we have narrated is an untested theory.
We leave it to future work to establish its empirical validity.

\subsection{Implications for Design}
By allowing users to create multiple communities with similar or identical topics, platforms can host ecosystems of online communities capable of providing a larger range of benefits to a larger range of users.
Some platforms, such as Stack Exchange, prohibit new communities from overlapping with existing communities \citep{fu_knowledge_2016}. 
Our findings suggest that such rules limit the range of benefits the platform's communities can confer.

Existing designs for online community platforms such as Reddit are at best ``first-order approximations'' of an ideal solution in that a ``sociotechnical gap'' remains between these designs and the goal of a platform that meets every person's every need \citep{ackerman_intellectual_2000}. Our interviewees partly filled this gap with personalized bespoke solutions in the form of their handpicked portfolios of communities. 
Improved designs for multicommunity discovery and engagement can better support users in knitting together these portfolios.

Many Reddit users make heavy use of the aggregated streaming feeds \texttt{r/all} and \texttt{r/popular}, which surface highly upvoted posts from across Reddit.  
Our interviewees described these feeds as most often featuring content from subreddits that are already extremely popular. Furthermore, Reddit's system for recommending subreddits often returned irrelevant suggestions.
Suggesting communities in as many cells in Figure \ref{fig:trilemma} as possible could help users build their portfolios of communities.
Because increased visibility may create stress and labor for communities and moderators \citep{kiene_surviving_2016}, recommendations should target those potential members likely to be positive contributors.

Although some of our interviewees used the ``multireddit'' feature for making a custom feed of subreddits, they described this feature as cumbersome and overwhelming.
A design alternative is to formalize or even automate the types of informal social practices our interviewees described such as cross-community linking and cross-posting. 
For example, a subreddit such as \texttt{r/vintageaudio} might configure an auto-moderator to detect posts about reel-to-reel equipment and recommend cross-posting to \texttt{r/reeltoreel}. 
A discussion-focused subreddit might routinely invite productive contributors to discussions in the related ``main'' subreddit.
Because intercommunity interactions can give rise to conflict, individual communities should have control of how such practices are implemented.
New tools for collaboration between moderation teams may enable the institution of policies encouraging productive concurrent participation in overlapping communities.

\subsection{Limitations}

Our study has limitations common to all interview-based studies. Our findings derive from in-depth conversations with relatively few of the people who were highly active participants in the handful of clusters of communities in our sample. 
Although our study was designed to achieve analytic saturation within each cluster and to cover a wide range of types of topics discussed on Reddit, additional interviews across a wider range of communities might uncover new types of specialization. Additionally, our interviewees were among the most active members of the clusters, and their experiences may differ from those of peripheral members.
Similarly, we cannot speak to the experiences of those who participated in only one community within a cluster.

Our interview data were collected at one point in time and cannot speak to how the dynamics we describe played out over time or how new communities were created and emerged.
Relatedly, although we find that overlapping communities tend to provide different benefits to members, we did not set out to interview community founders and thus cannot speak to the reasons that communities were created \cite{foote_starting_2017}. 

Furthermore, our study focuses only on the Reddit platform.   Reddit has distinctive affordances for voting, moderation, and multicommunity engagement that might shape the construction and use of overlapping communities. Although Reddit is among the most popular online community platforms. Our findings may not describe relationships between overlapping communities on other platforms, or between one platform and another. Different platforms likely have different strengths or weaknesses for building communities that provide some types of benefits but not others. 
At the same time, cross-platform engagement may involve friction related to the use of multiple identities and sociotechnical systems.
Future research should investigate how people use portfolios that include communities on multiple platforms.
 
\section{Conclusion}
Why are the same people talking to each other about similar things in different online communities? We answer this question by developing a theory grounded in the analysis of 20 interviews with members of highly related communities on Reddit. Our answer suggests that people turn to online communities in search of multiple benefits---specific kinds of content and discussion, socialization in a homophilous community, and attention from the largest possible audience. We argue that although structures such as the topic, rules, and size of a community might improve the degree to which it provides one of these benefits, they will necessarily detract from its ability to provide others.  Multiple communities having a range of structures exist to provide the full range of benefits. No community can do everything.

\begin{acks}
Text from a draft of this article was included as part in the first authors PhD dissertation and he thanks his committee---Professors Aaron Shaw, Emma Spiro, David McDonald and Kirsten Foot---for their generous support, wise advice and insightful comments. This work was supported by NSF grants IIS-1908850, IIS-1910202, and GRFP \#2016220885. Versions of this paper received very helpful feedback from the Community Data Science Collective and Aaron Shaw, Sohyeon Hwang, Jeremy Foote,  Carl Colglazier, Floor Fiers, Sejal Khatri, Sefania Druga, Nicholas Vincent, and Kaylea Champion in particular. We also thank the HCI seminar at the University of Washington for their thoughtful discussion and comments. 
We are grateful to the anonymous CSCW reveiwers for their keen insights and feedback which considerably improved the paper.
Thanks to Jason Baumgartner and pushshift.io for the Reddit data archive. This work was facilitated through the use of the advanced computational infrastructure provided by the Hyak supercomputer system at the University of Washington. 
Most of all, we owe special gratitude to our 20 interview participants for their time and knowledge.
\end{acks}

\bibliographystyle{ACM-Reference-Format}
\bibliography{references}
\end{document}